 \documentclass{elsart}

\def\de{\delta}
\def\ep{\epsilon}
\def\la{\lambda}

\def\krn{k}

\def\pl{\partial}
\def\cd{\cdot}
\def\spa{\ \ \ }
\def\mfor{\spa\mbox{for}\spa}
\def\main{\spa\mbox{in}\spa}
\def\maon{\spa\mbox{on}\spa}
\def\mwhere{\spa\mbox{where}\spa}
\def\ha{\mbox{$\frac{1}{2}$}}
\def\vmax{V_{\mathrm{max}}}

 \usepackage{graphicx}

\usepackage{amssymb}

\usepackage{psfrag}
\usepackage{cite}

\begin{document}

\begin{frontmatter}



\title{Fracture resistance via topology optimisation}


\author[UQ]{V J Challis}
\ead{vchallis@maths.uq.edu.au}
\author[UQ]{A P Roberts}
\ead{apr@maths.uq.edu.au}
\author[UQ]{A H Wilkins\corauthref{andy}}
\ead{awilkins@maths.uq.edu.au}

\corauth[andy]{Corresponding author.  Telephone: +61 7 3365 3266.
  Fax: +61 7 3365 1477}

\address[UQ]{Department of Mathematics, University of Queensland, Brisbane, QLD 4072, Australia}

\begin{abstract}
The fracture resistance of structures is optimised using
the level-set method.  Fracture resistance is assumed to be related to
the elastic energy released by a crack propagating in a normal
direction from parts of the boundary which are in tension, and is
calculated using the virtual crack extension technique.  The shape
derivative of the fracture-resistance objective function is derived.
Two illustrative two-dimensional case studies are presented: a hole
in a plate subjected to biaxial strain; and a bridge fixed at both
ends subjected to a single load in which the compliance and fracture
resistance are jointly optimised.  The structures obtained have
rounded corners and more material at places where they are in tension.
Based on the results, we propose that fracture resistance may be
modelled more easily but less directly by including a term
proportional to surface area in the objective function, in conjunction
with non-linear elasticity where the Young's modulus in tension is
lower than in compression.
\end{abstract}


\end{frontmatter}

\section{Introduction}
\label{intro}
Topology optimisation involves finding the geometry of an object which
minimises an objective function, with no constraints on the object's
topology~\cite{bendsoeET04}.  Broadly speaking, there are currently two main methods
for studying such problems: the homogenisation method and its variants
such as the well-established SIMP method~\cite{bendsoeET88,bendsoe89};
and, the level-set method~\cite{osherET88}.  In the former, the
optimisation proceeds via a sequence of `fictitious' materials which
have properties unrealizable in nature until converging to the minimum
containing only real
materials~\cite{suzukiET91,allaireET97,bendsoeET99,allaire01}.  In the
latter, the process uses only real materials and moves boundaries to
decrease the objective function at every
iteration~\cite{sethianET00,osherET01,allaireET04,wangET03}.  In its
most naive implementation it typically converges slowly towards a {\em
local} minimum, although recently some authors have proposed more
sophisticated strategies to address these
problems~\cite{burgerET04,wangET04,allaireET05a,hintermuellerET05,guoET05,amstutzET06,degournay06,gomesET06,wangET05c,wangET06c,wang06,wilkeET06}.
These strategies are not used here.

Topology optimisation has been used to study many problems, 
and in particular the elastic problem, where the objective
function is the object's compliance or total elastic energy for a
particular set of external loads and boundary constraints, has been
well studied.  Bends{\o}e {\em et al.}~\cite{bendsoeET05}
have prepared a recent review which discusses problems involving
pressure loads, electromagnetic and acoustic wave propagation, fluid
flow, articulated mechanisms and buckling.  

Level-set methods were first applied to structural optimisation in 
\cite{sethianET00} and have since been applied to problems in two and three dimensions including compliance minimisation with single and multiple load cases 
~\cite{allaireET04,wangET03,wangET04,allaireET05a,guoET05,amstutzET06,degournay06,gomesET06,wangET05c,wangET06c,wang06,allaireET02,yulinET04,wangET04a,allaireET05b,wangET05b} , 
multi-material problems~\cite{yulinET04,wangET04a,wangET05b,wangET05},
eigenfrequency problems~\cite{osherET01,degournay06,allaireET05b}, gripping and
clamping devices~\cite{allaireET04,allaireET05a,amstutzET06,degournay06,allaireET02,yulinET04,wangET05}, and shape and image reconstruction~\cite{burgerET04,hintermuellerET05}.

The problem of optimising a material's fracture toughness is addressed
in this paper.  Our definition of fracture toughness involves
consideration of cracks propagating in a normal direction from any
point on the object's boundary into its interior.  Thus, the level-set
method is used since the objects of the intermediate iterations in the
homogenisation/SIMP method have no well-defined ``boundaries'', being
solid masses of fictitious materials.  The use of this method also
removes the problem of defining the fracture toughness of
fictitious materials.

The examples presented are two dimensional (2D).  They are biphasic,
consisting of ``solid'' which has a Young's modulus of unity and a
Poisson's ratio of 0.3, and ``void'' which has a Young's modulus of
zero.

Our approach closely follows Allaire {\em et al.} and for brevity we
do not reproduce their methodology, rather we refer readers to the
article~\cite{allaireET04}. 

\newpage

\section{Brittle fracture and the objective function}

According to the classic work of Griffith~\cite{griffith21}, a brittle
material fractures if the elastic energy released upon crack
propagation exceeds the material's ``fracture surface energy'', which
is a material-specific quantity parameterising the strength of the
intermolecular bonds~\cite{lawn93}.  Given this quantity, the maximum
allowed load for a particular situation may be obtained by calculating
the elastic energy released for all possible crack propagation paths.
A crack can nucleate and propagate from within a material, or it may
nucleate from its boundary and propagate inwards.  The latter
situation only is considered here.

The energy release rate for crack propagation may be calculated by
finding the material's elastic energy for two configurations
differing only by a small change in crack length.  In practice, where
the two energies are calculated numerically, this is computationally
costly, and for the linear elastic case methods exist which only
require one elastic relaxation, such as (for 2D) the
J-integral~\cite{rice68}, or equivalently, the virtual crack extension
method~\cite{parks74,hellen75}.

Special cases of these methods may be interpreted as a particular
shape derivatives as follows.  Denote the domain occupied by the
elastic material by $\Omega$.  It has boundary $\pl\Omega =
\Gamma_{N}\cup\Gamma_{D}$ with outward unit normal $n$.  The material is
subjected to external tractions $g$ on $\Gamma_{N}$ (Neumann
conditions), body forces $f$ in its interior, and its boundary has fixed
displacement, $u^{D}$,  on $\Gamma_{D}$ (Dirichlet conditions).  The
displacement field that solves this linear elastic problem is denoted
by $u_{i}$, and denote the energy density by $W$:
\[
W = \ha\nabla_{i}u_{j}A_{ijkl}\nabla_{k}u_{l} \ .
\]
Here and below the indices $i,j,k,l = (1,\ldots d)$ in $d$ dimensions,
summation over repeated indices is implied, the index $n$ is reserved for
the ``normal direction'' (e.g. $x_{n}=x_{i}n_{i}$ for any vector $x$), and
$\nabla_{i}$ is the 
derivative, so the strain is \mbox{$\ha(\nabla_{i}u_{j}+\nabla_{j}u_{i})$}.
The elastic tensor is denoted by $A_{ijkl}$.   The compliance is
\begin{equation}
C = \int_{\Omega}f_{i}u_{i} + \int_{\Gamma_{N}}g_{i}u_{i} \ .
\end{equation}

Consider deforming the 
material (coordinatised by $x$) with the diffeomorphism
\[
x\rightarrow x+\theta(x) \ .
\]
Classical
results~\cite{muratET76,simon80,sokolowskiET92,allaireET04} give the
compliance change as a function of $\theta$ as
\begin{eqnarray}
C'(\theta) & = & 2\int_{\Gamma_{N}}\theta\cd
n\,\left(\nabla_{n}(g\cd u) + H g\cd u - W\right)
\nonumber \\
&& \spa + 2\int_{\Gamma_{D}}\theta\cd
n\,\left(\nabla_{n}(u_{j}-u^{D}_{j})A_{njkl}\nabla_{k}u_{l} - W\right) \ .
\label{gen.com.ch.}
\end{eqnarray}
In this formula $\nabla_{n}$ denotes the derivative in the normal
direction and $H=\nabla \cd n$ is the mean curvature of the boundary.  

\begin{figure}[htb]
\centering
\psfrag{y=0}{$y=0$}
\psfrag{y=d}{$y=\de$}
\psfrag{y=-d}{$y=-\de$}
\psfrag{e}{$\ep$}
\psfrag{O}{$\Omega$}
\psfrag{gnugd}{$\Gamma_{N}\cup\Gamma_{D}$}
\includegraphics[width=10cm]{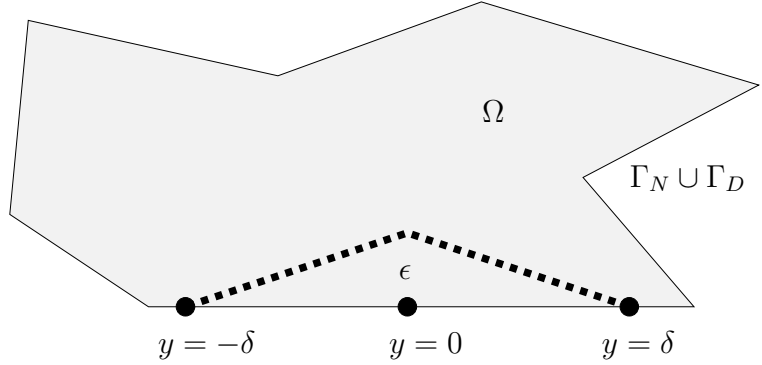}
\caption{The perturbation of the boundary given by Eq.~(\ref{def.of.f.eqn}).}
\label{krn.fig}
\end{figure}

To make the connection with the J-integral and virtual crack
extension, consider the two dimensional (2D) case.
Parameterise length along the boundary by $y$.  (The boundary may
consist of disjoint pieces, but that is irrelevant because of the
following assumption.)  Assume that $\theta$ takes the form
\begin{equation}
-\theta\cd n = \ep\krn_{0}(y) = \left\{
\begin{array}{ll}
\ep (\de+y) & \mfor -\de\leq y\leq 0 \\
\ep(\de-y) & \mfor 0\leq y\leq \de \\
0 & \spa\mbox{otherwise}
\end{array}
\right\}\spa\mbox{on }\pl\Omega.
\label{def.of.f.eqn}
\end{equation}
Here $\ep$ is a small number parameterising the ``depth'' of the
boundary perturbation, and $\de$ parameterises the ``length'' of the
perturbation with $\de\gg\ep$.  The perturbation is shown in
Fig.~\ref{krn.fig}.
The subscript on $\krn$ indicates that the
perturbation is rooted at $y=0$, and suppose that $y=0$ is part of
$\Gamma_{N}$.  This is exactly a particular
``virtual crack extension'':  in a finite element scheme the point $y=0$ is
one node of the finite element mesh, $\krn_{0}$ corresponds to the
virtual movement of this single node while keeping all other nodes fixed.  
Specialising Eq.~(\ref{gen.com.ch.}) to this case gives
\begin{equation}
C'(\theta) = \int\d y\ \ep\krn_{0}(y)(W-\nabla_{n}(g\cd u) - Hg\cd u) \
.
\label{v.c.e.f}
\end{equation}
When the section of boundary $-\de\leq y\leq \de$ is straight, $H=0$,
and this expression may be compared with Rice's J-integral evaluated
along a contour close to the point $y=0$.

In standard fracture situations where the material contains a
well-defined crack of interest, numerical errors in the elasticity
solution mean it is more accurate to move all the nodes which lie in
the vicinity of a crack tip --- $\krn$ has nonzero support at all of
these nodes --- in order to calculate the energy release rate (or
compliance change) for crack propagation in a certain direction.
Similarly, for well-defined straight cracks, errors can be minimised
by evaluating the J-integral around a contour that is distant from the
crack tip.  In the current application, however, generically there
will be no well-defined cracks, and so the movement of just one node
is considered, corresponding to a small crack initiating and
propagating from the boundary in the normal direction.


Motivated by these arguments, the objective function for fracture
resistance considered here is
\begin{equation}
J_{q} = \int_{p\in S_{+}}\krn_{p}\,G^{q} \mwhere
G=W-\nabla_{n}(g\cd u)-Hg\cd u \ .
\label{obj.fract}
\end{equation}
The short-hand notation, $G$, has been introduced.    The set
$S_{+}$ is all the points on $\Gamma_{N}$ where the surface tension is
positive, so as to exclude points of high compressive energies
which may not actually lead to fracture.  Further comments on this
point are made in Sec.~\ref{sec.discussion}.  An explicit restriction to
$\Gamma_{N}$ has been made, since the material is fixed at
$\Gamma_{D}$ so that part of the boundary cannot be moved, so its
fracture resistance cannot be optimised.  When the exponent, $q$, is
large, minimising $J_{q}$ corresponds to focussing on those few boundary
points where the fracture resistance is low, while smaller $q$ (say
$q=1$) considers points in $S_{+}$ more equally.  The latter approach
has been found to lead to more rapid convergence of the numerical
algorithm, since the `special points' in the former tend to move
around during optimisation.

\section{The shape derivative of the objective function}

The shape derivative of the objective function is needed to implement
the level-set scheme as detailed in Allaire {\em et
  al.}~\cite{allaireET04}, since the velocity in the Hamilton-Jacobi (HJ)
equation is the negative of the integrand of the shape derivative.

To find the shape derivative of Eq.~(\ref{obj.fract}), the formal method
discussed in~\cite{allaireET04} and attributed to
C{\'{e}}a~\cite{cea86} is used.  The idea is to find a Lagrangian, $L$,
which generates the equations of motion for the field(s) and reduces
to the objective function when the fields satisfy their equations of
motion (``on shell'').  Because of the former property, the derivative $\de
L/\de\mbox{fields}=0$ on shell, so the shape derivative $dL/d\Omega$
can be computed at fixed fields.

The elasticity problem is defined through the equations of motion
\begin{eqnarray}
\nabla_{j}(A_{ijkl}\nabla_{k}u_{l}) & = &  -f_{i} \main \Omega \ ,
\nonumber \\
A_{njkl}\nabla_{k}u_{l} & = & g_{j} \maon \Gamma_{N} \ , \nonumber \\
u_{i} & = &  u_{i}^{D}\maon \Gamma_{D} \ .
\label{gen.eoms.u}
\end{eqnarray}
Written in terms of stress, $\sigma$, the first is simply 
$\nabla_{j}\sigma_{ij}=-f_{i}$ and the second is $\sigma_{nj}=g_{j}$,
so $f$ is a body force and $g$ a boundary traction.

There are many Lagrangians which generate these equations of motion,
and the following contains just one auxialiary field $M$ which is a
Lagrange multiplier:
\begin{eqnarray}
L_{M} & = & \int_{\Omega}\nabla_{i}M_{j}A_{ijkl}\nabla_{k}u_{l} -
\int_{\Omega}M_{i}f_{i} - \int_{\Gamma_{N}}M_{i}g_{i} \nonumber \\
&& - \int_{\Gamma^{D}}\left[M_{j}A_{njkl}\nabla_{k}u_{l} +
(u_{j}-u_{j}^{D})A_{njkl}\nabla_{k}M_{l} \right] \ .
\label{eqn.l.m.2}
\end{eqnarray}
Varying $M_{i}$ gives the correct equations of motion for $u$; and
\[
L_{M}=0 \ ,
\]
on shell.

Therefore, the Lagrangian
\[
L = J_{q} + L_{M} \ ,
\]
satisfies the requirements given above.  Varying $u$ in this
Lagrangian gives the equations of motion for the auxiliary field:
\begin{eqnarray}
\nabla_{j}\left(A_{ijkl}\nabla_{k}M_{l}\right) & = & 0 \main \Omega \
, \nonumber \\ 
A_{njkl}\nabla_{k}M_{l} & = & 
\nabla^{t}_{i}\left(\krn_p qG^{q-1}A_{ijkl}\nabla_{k}u_{l}
\right) + \krn_{p} qG^{q-1}\left(\nabla_{n}g_{j} + Hg_{j}\right) \maon
\Gamma_{N} \nonumber \ , \\  
M_{i} & = & 0 \maon \Gamma_{D} \ .
\label{eoms.al.bdy}
\end{eqnarray}
The derivative $\nabla^{t}_{i}$ acts in the
tangent plane only:
\[
\nabla^{t}_{i} = \nabla_{i}-n_{i}\nabla_{n} \ .
\]
In deriving these equations, the boundary conditions on $u$ have been
used, and it is assumed that $\krn_{p}$ has zero support on
$\pl\Gamma_{N}$, so that there is no boundary term when integrating by
parts with the operator $\nabla^{t}_{i}$.  These equations therefore
define an elasticity problem with zero displacement on $\Gamma_{D}$
and tractions applied to the boundary around the points $p\in S_{+}$.

Using the standard formulae found in~\cite{allaireET04}, and
evaluating on-shell (so that $u=u^{D}$ and $M=0$ on $\Gamma_{D}$, for
instance), the shape derivative of the objective function is
\begin{eqnarray}
J_{q}'(\theta) & = &
\int_{S_{+}}
\left(\nabla_{n}\left(\krn_{p}G^{q}\right) + H\krn_{p}G^{q} \right)\,\theta\cd n 
+ \int_{\pl\Omega}\nabla_{i}M_{j}A_{ijkl}\nabla_{k}u_{l}\ \theta\cd n
\nonumber \\
&& - \int_{\Gamma_{N}}\left(
\nabla_{n}(g\cd M)+Hg\cd M\right)\,\theta\cd n
\nonumber \\
&& - \int_{\Gamma_{D}}\left(
\nabla_{n}M_{j}A_{njkl}\nabla_{k}u_{l} +
\nabla_{n}u_{j}A_{njkl}\nabla_{k}M_{l} \right)\,\theta\cd n
.
\label{eqn.shape.deriv.frac.}
\end{eqnarray}

\section{Compliance minimisation and the volume or area constraint}

In addition to fracture resistance, compliance minimisation is also
considered in the examples below.  The objective function used is
\begin{equation}
J = (1-\la)C + \la J_{q} \ ,
\end{equation}
where $\la$ is a fixed weighting factor.  Both $C$ and $J_{q}$ are
minimised when the material has infinite volume (3D) or area (2D).  To
avoid this, a maximum volume (3D) or area (2D), $\vmax$, is introduced, and the
problem to be solved is  
\begin{equation}
\mbox{Minimise $J$ subject to }\int_{\Omega} 1 \leq \vmax \ .
\label{prob.stated}
\end{equation}
It is common in the literature to add a term proportional to the
volume (or area) to the objective function, so the problem becomes:
minimise a linear combination of $J$ and volume (or area).  Unfortunately
this does not allow direct comparison of structures with different
$\la$ since altering $\la$ will also produce different volume
fractions.  The problem stated in Eq.~(\ref{prob.stated}) does allow a
direct comparison.

\section{Finite element implementation}

\begin{figure}[htb]
\centering
\psfrag{e}{$\ep$}
\psfrag{e2}{$\frac{\ep}{\sqrt{2}}$}
\psfrag{dd}{$\de$}
\psfrag{x}{$x$}
\psfrag{y}{$y$}
\includegraphics[width=14cm]{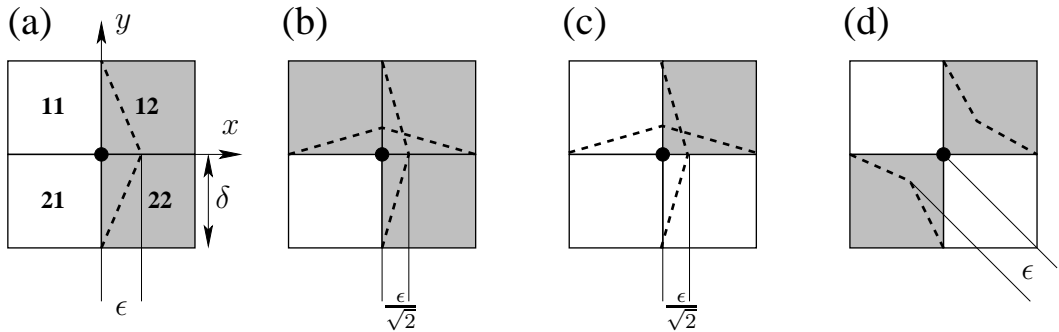}
\caption{Four possibilities for a node (black dot) on the boundary.
  (a): there are two neighbouring solid elements and two void
  elements.  (b): three solid elements.  (c): one solid element.  (d):
  two diagonally-opposite solid elements.  Virtual crack extensions
  for each are shown by dotted lines: (b), (c) and (d) can be
  decomposed into a linear combination of two orthogonal extensions.
  The surrounding elements are labelled 11, 12, 21 and 22 as shown in (a).}
\label{fig.4.options}
\end{figure}

The finite element method is implemented using bi-linear (in 2D)
square elements, as described
in~\cite{garbocziET95,garboczi98,wilkins06}.  The original ($u$)
elastic problem is solved, and $G$ at each boundary node is
calculated to find the boundary tractions for the auxiliary ($M$)
problem.  Four geometric arrangements are possible, as
illustrated in Fig.~\ref{fig.4.options}.  In the case of
Fig.~\ref{fig.4.options}(a), the integral becomes
\begin{equation}
\int \krn_{p}G^{q} = \int_{-\de}^{0}\d y\ \krn_{0}
(G_{22}^{q}-G_{21}^{q})_{x=0} + \int_{0}^{\de}\d y\ 
\krn_{0}(G_{12}^{q}-G_{11}^{q})_{x=0} \ ,
\label{explicit.j.calc}
\end{equation}
where the subscripts 11, etc, correspond to the values in the elements
defined in Fig.~\ref{fig.4.options}.  In calculating these $G$ values,
$H=0$ since the boundary is straight, and in the case illustrated
$G_{11}=0=G_{21}$.  In principle, if there is a boundary
traction, $g_{i}$, acting at the central node then its action gets split
over the elements 12 and 22 at $y=0$ (however, in the practical
situations considered here and elsewhere the boundary never moves away
from such a point so the point may be simply excluded from $S_{+}$).

In the case of Fig.~\ref{fig.4.options}(b) and (c)
the normal is taken to be $(-1,-1)/\sqrt{2}$ to ascertain whether the
point is in $S_{+}$.  Algorithmically it is convenient to use the sum
of two orthogonal virtual crack extensions as indicated in the figure,
each with a  similar expression to Eq.~(\ref{explicit.j.calc}).  In the case of
Fig.~\ref{fig.4.options}(d), the normal direction is not well-defined.
In this case, the two normal virtual crack extensions are tried (each
can be decomposed into two orthogonal parts as in (b) and (c)), and the
maximum of the two results (which will be numerically similar) is
chosen.

The elasticity problem is then solved for $M$.  The surface tractions
are applied at each point $p\in S_{+}$ and its two neighbouring nodes on the
boundary.  A finite difference scheme is adopted to find the
tangential divergence in Eq.~(\ref{eoms.al.bdy}):
$k_{p}qG^{q-1}\sigma_{ij}$ is evaluated at points half way between the
nodes ($y=\pm\de/2$ with $x=0$ in
Fig.~\ref{fig.4.options}(a)) and the appropriate differences taken.
Cases Fig.~\ref{fig.4.options}(b), (c) and (d) follow similarly, for
instance, in case (b)
\begin{eqnarray}
\left.\nabla^{t}_{i}k_{0}qG^{q-1}\sigma_{ij}\right|_{x=0=y} & = &
 \frac{1}{\de / \sqrt{2}}\left(\left. k_{0}qG^{q-1}\frac{(\sigma_{1j}-\sigma_{2j})}{\sqrt{2}}\right|_{x=-\de/2,y=0} \right. \nonumber \\
& & \left.- \left. k_{0}qG^{q-1}\frac{(\sigma_{1j}-\sigma_{2j})}{\sqrt{2}}\right|_{x=0,y=-\de/2} \right)\nonumber
\end{eqnarray}
since the tangential stress is
$(\sigma_{1j}-\sigma_{2j})/\sqrt{2}$ and the distance between
the points $(-\de/2,0)$ and $(0,-\de/2)$ is $\de/\sqrt{2}$.

After the elasticity problem for $M$ has been solved, the shape
derivative of $J_{q}$ can be found by straightforward evaluation.  The
only nontrivial term in Eq~(\ref{eqn.shape.deriv.frac.}) is
$\nabla_{n}\left(k_{p}G^{q}\right)$.  For Fig.~\ref{fig.4.options}(a), this is
evaluated by by taking the finite difference of
Eq.~(\ref{explicit.j.calc}) at $x=1$ (using a Young's modulus of zero for 
the 12 and 22 elements in the virtual crack extension calculation) with
that at $x=0$.  Similar constructions are used for the other 3 cases
in Fig.~\ref{fig.4.options}. 

The void material has zero Young's modulus and thus the integrand of
the shape derivative is zero for void elements.  Consequently, the
velocity, $v$, of the HJ evolution is zero too, and so the
boundaries of the object will never move!  Allaire {\em et
al.}~\cite{allaireET04} promote the use of a so-called ersatz
material, where the voids have a small nonzero Young's modulus,
allowing the velocity to be extended into the void elements.  This is
not suitable here, since the first and third terms of
Eq.~(\ref{eqn.shape.deriv.frac.}) are only {\em defined} on the boundary.
Therefore, a smoothing operation is performed as follows.  The first
and third terms are defined at the {\em nodal} points, so their
contribution is spread equally to the four neighbouring elements.  The
second is evaluated at the midpoints of each of the solid elements, and then spread to
the neighbours by the correlation:
\[
v_{\mathrm{new}}(i,j) = (2v(i,j)+v(i-1,j)+v(i+1,j)+v(i,j-1)+v(i,j+1))/6 \ .
\]
This smoothing of the boundary values into the void allows the
boundary to move during HJ evolution.

As found by other authors, the objective function typically decreases
with each HJ iteration, only increasing when the solid volume fraction
exceeds $\vmax$.  When that occurs, the material is eroded away using
a constant velocity in the HJ evolution until the volume fraction is
$0.99\vmax$.  The behaviour of $J_{q}$ is not as controlled as in the
compliance-minimisation case because sometimes the movement of one
part of the boundary causes $G$ to increase markedly somewhere else.
This may be seen in Fig.~\ref{j.eg}.  Because our finite-element
program is parallelised and hence rather quick, we evolve very slowly:
each iteration involves an elastic relaxation and movement of the
boundary by only {\em one} element.

\section{Case study 1: the hole}

\begin{figure}[htb]
\centering
\includegraphics[width=14cm]{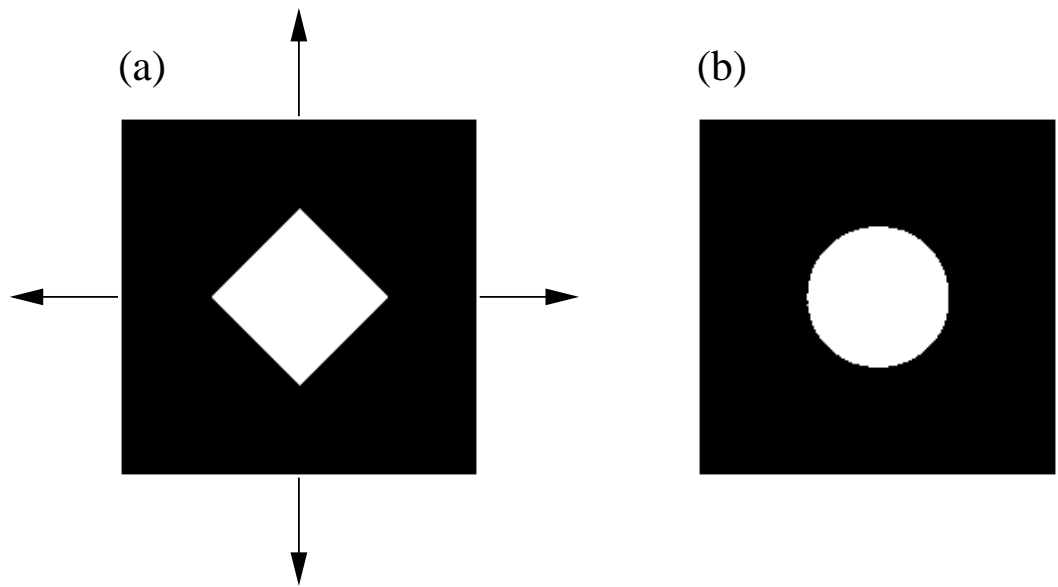}
\caption{(a): An initial configuration with $J_{2}=973$.  (b) The
  minimum at $J_{2}=177$.  This example has $200 \times 200$ pixels.}
\label{hole.fig}
\end{figure}

The fracture resistance of a plate containing an arbitrarily shaped
central hole must be maximised under the conditions that the hole area
fraction is $1/8$ and the plate is under uniform biaxial strain.  This
problem may be solved using $\la=1$ and $\vmax=7/8$.
Fig.~\ref{hole.fig} shows one initial condition, and the geometry at
the minimum, which has a circular hole.  Because this is the expected
result, this case study illustrates that the algorithm is indeed
maximising fracture resistance.

\section{Case study 2: the bridge}

\begin{figure}[p]
\centering
\includegraphics[width=14cm]{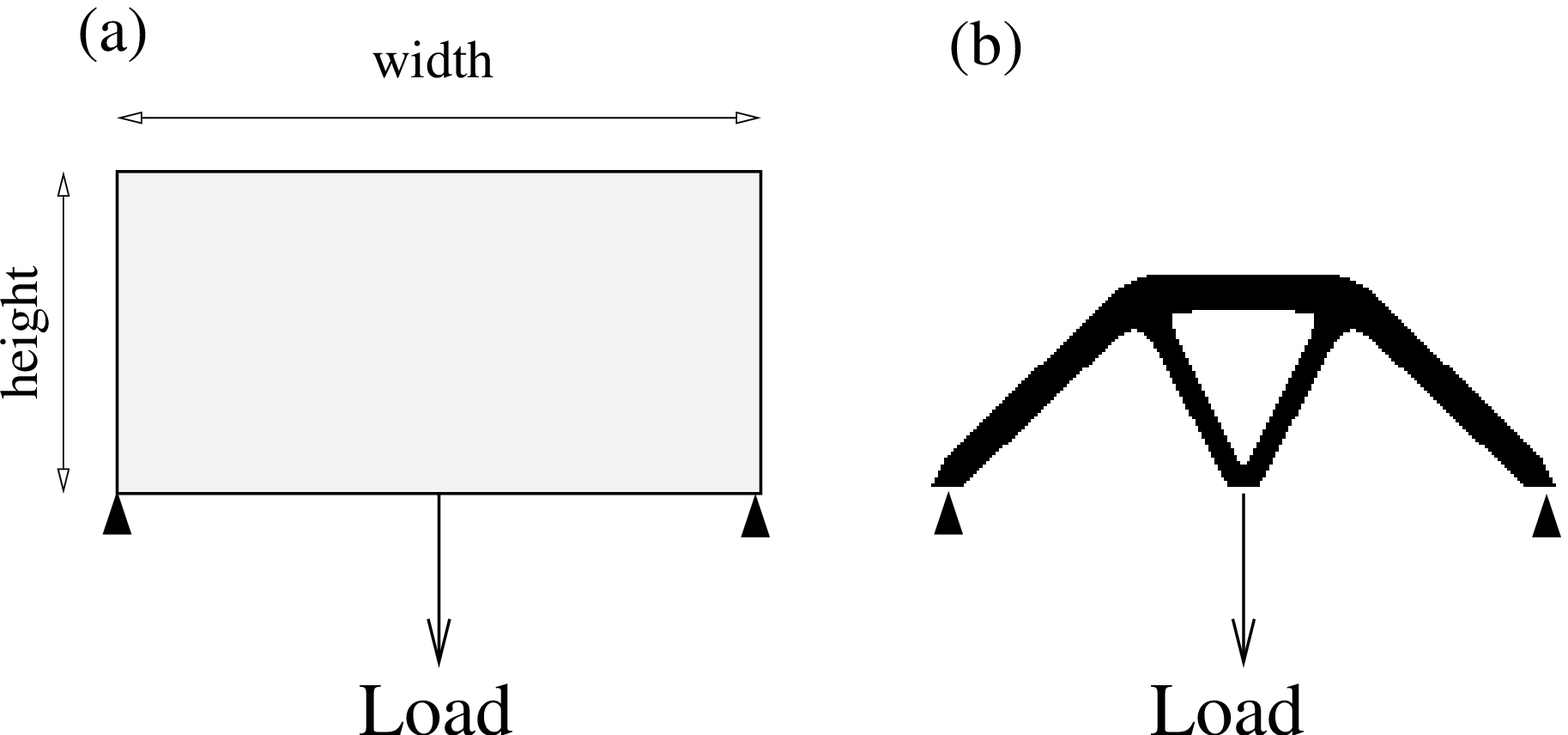}
\caption{(a): The bridge problem has two points with fixed zero
  displacement and a single load.  (b): The solution for
  $\mbox{width}=2\times\mbox{height}$, $\vmax=17.1$\% and $\la=0$.}
\label{bridge.f0.fig}
\end{figure}

The `bridge' problem is specified in Fig.~\ref{bridge.f0.fig} which
also shows its solution for $\la=0$, $\vmax=17.1$\%, and the length
variable being twice the height.  Similar problems were studied
in~\cite{bendsoeET04,allaireET04,wangET03,wangET04,allaireET05a,amstutzET06,wangET05c,wangET06c}.  The
compliance of the structure shown in Fig.~\ref{bridge.f0.fig}(b) is
$C=26.5$.  This topology and geometry is also obtained using the SIMP
method~\cite{bendsoeET04}.  The simulations were performed on a
$200\times 100$ mesh.  A large variety of initial configurations were
used in order to address the problem of converging to a local minimum,
rather than a global one.

The minimum structure at $\la=0$ has $J_{1}=0.5$ --- 50 times less
than $C$ --- so to weight fracture resistance meaningfully a value of
$\la\approx 1$ must be used.  The following study considers the
extreme case of $\la=1-10^{-5}$.  Less extreme values produce
geometries that interpolate between the one shown here and the $\la=0$
case.  In order to keep the structure within the rectangular domain of
interest, it is assumed that any parts of the bridge on the bottom
boundary are infinitely fracture resistant, otherwise the optimisation
results in material enveloping the point of application
of the load.

\begin{figure}[p]
\centering
\begin{tabular}{cc}
\scalebox{1.0}{\includegraphics{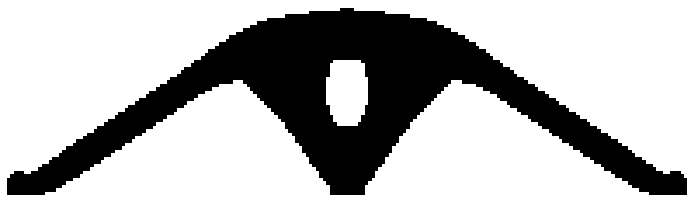}} &
\scalebox{0.25}{\includegraphics{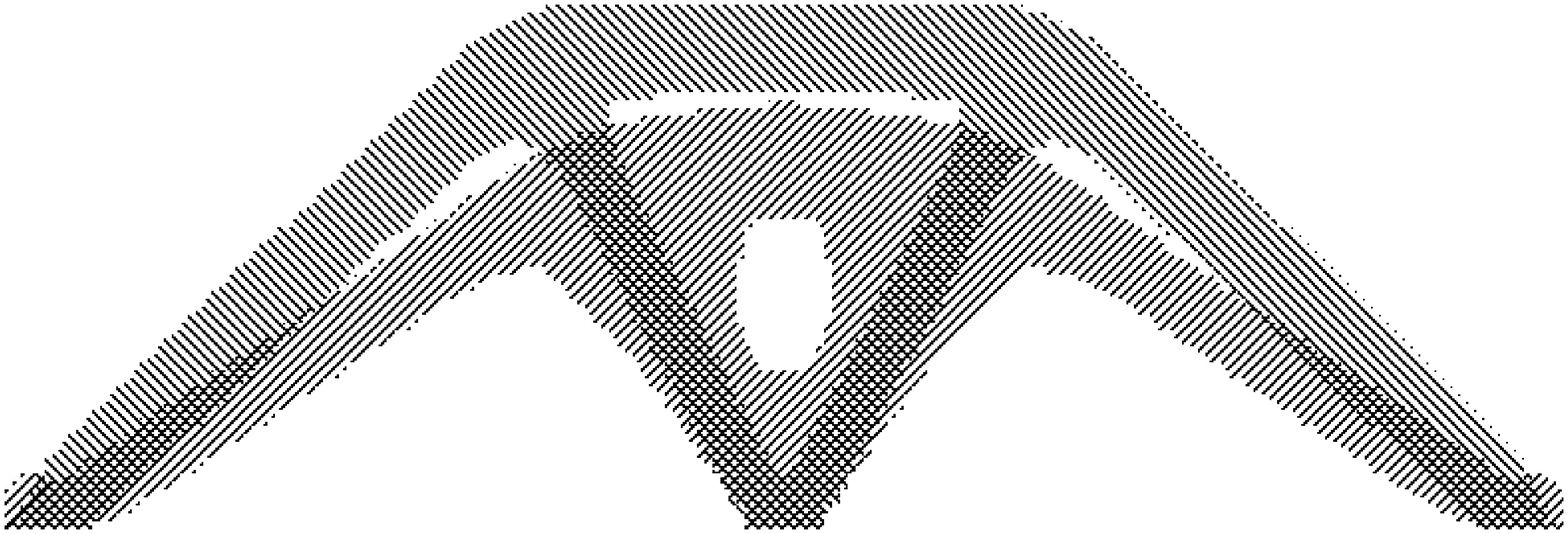}}
\end{tabular}
\caption{Left: The optimal geometry for the bridge problem for $\la=1-10^{-6}$.  Right: a
  comparison of the left-hand structure and the optimal geometry for $\la=0$.}
\label{frac.eg}
\end{figure}

\begin{figure}[p]
\centering
\scalebox{0.5}{\includegraphics{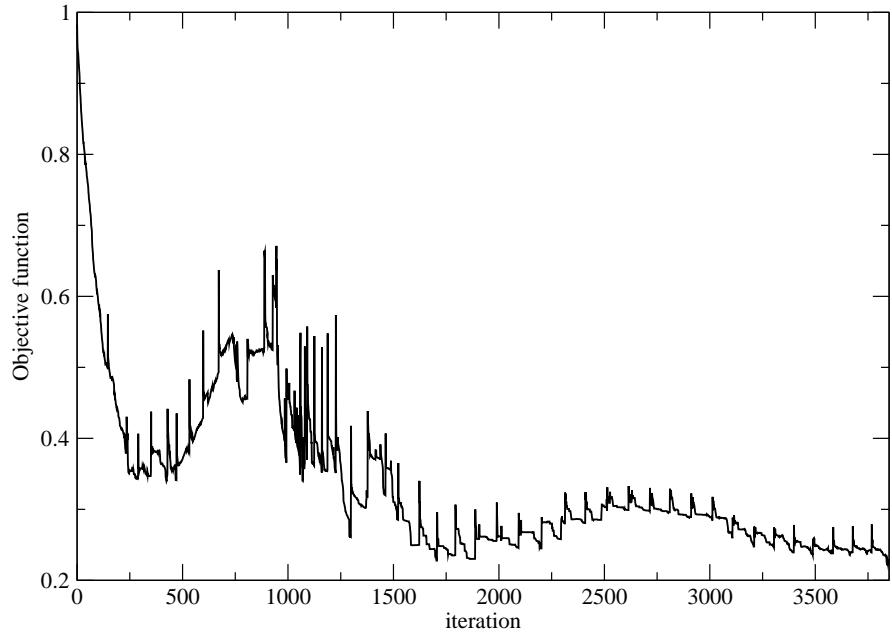}}
\caption{The objective function at each iteration for the bridge case study.
  Each iteration involves either: adding {\em two} elements
  (symmetrically about the symmetry axis) to the solid structure; or,
  eroding the structure so that the solid volume fraction is
  $0.99\vmax$.  The latter causes large jumps while the former typically
  decreases $J$, as discussed in the text.  The initial configuration
  had topology and geometry similar to that of the
  compliance-optimised solution of Fig.~\ref{bridge.f0.fig}(b).}
\label{j.eg}
\end{figure}

\begin{figure}[p]
\centering
\scalebox{0.9}{\includegraphics{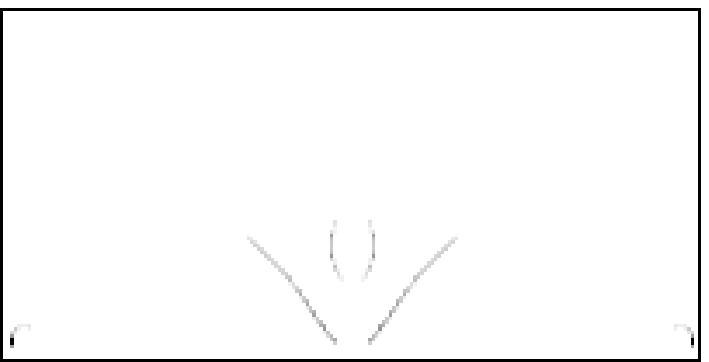}} 
\caption{Greyscaling indicating the value of $G$ for each boundary
  node in the optimal solution shown in Fig.~\ref{frac.eg}.  Dark
  regions, such as those close to the supporting points indicate high
  $G$, lighter regions indicate lower $G$. }
\label{g.opt}
\end{figure}

Figure~\ref{frac.eg} shows the optimal geometry, and Fig.~\ref{j.eg}
shows the objective function.  This geometry has $C=34.3$,
$J_{1}=0.222$ and so $J=0.222$.  The following points may be noticed.
\begin{enumerate}
\item The parts of the structure which are in tension capture material
  from those in compression.
\item Corners are typically more rounded.
\item In the case of the bridge, the optimal topology is unchanged.
  In other cases that have been investigated (such as the cantilever),
  this is sometimes not true since a thin strut in compression can
  disappear entirely.
\item $G$ is fairly constant over the parts of the boundary that are
  in tension, as shown in Fig.~\ref{g.opt}.  This suggests that very
  similar geometries are obtained for $q>1$ (although the value of
  $\la$ must also be scaled to obtain exactly the same geometry).
\end{enumerate}
These points are supported by further simulations with different $q$
and $\la$: because the structures obtained are actually quite similar
we resist reproducing further graphics for brevity.

\section{Discussion} \label{sec.discussion}

In this paper fracture resistance has been defined through a virtual
crack extension in a normal direction on parts of the boundary that are
in tension.  Because the normal stresses are typically close to zero
on the boundary (even considering the few places where $g\neq 0$,
discretisation effects and numerical errors which all yield nonzero
normal stresses), the objective function for fracture resistance is
close to an integral of the energy density over parts of the boundary,
$S_{+}$, which are in tension.

If this restriction to $S_{+}$ was removed, we propose that the the
objective function for fracture resistance may be well approximated by
a {\em surface integral} of the energy density, and therefore, when
used in linear combination with the objective function for compliance,
which is an integral of the energy density over the {\em volume} of
the material, may be further replaced by an objective function
proportional to the {\em surface area} (in 3D) or {\em perimeter} (in
2D).  This provides a physical motivation for using a surface
area or perimeter constraint/objective function, which are typically
employed to penalise micro-fine structures sometimes encountered in
structural optimisation problems.

The problem is more complicated with the restriction to $S_{+}$, as
material is captured by the parts in tension.  Those parts do not
necessarily decrease significantly in surface area, but because they
are more massive will have significantly less surface energy.  Our
experience is that in most cases the optimal topology does not change
compared with the compliance-only situation, although the geometry may
be significantly altered, as can be seen in Fig.~\ref{frac.eg}.
Similarly to above, we propose that this case may be modelled less
directly, but probably more easily by including a term proportional to
the surface area or perimeter in the objective function, in conjunction
with using a non-linear elasticity law where the Young's modulus in
tension is lower than in compression.

\section{Acknowledgements}
This work was supported by a grant from the Australian Research
 Council through the Discovery Grant scheme, and an Australian
 Postgraduate Award.

\bibliographystyle{elsart-num}
\bibliography{refs}

\end{document}